# Wafer-Scale and Deterministic Patterned Growth of Monolayer MoS$_2$ via Vapor-Liquid-Solid Method

*Shisheng Li,\*[1,2] Yung-Chang Lin,[3] Xu-Ying Liu,[1,4,5] Zehua Hu,[6] Jing Wu,[7] Hideaki Nakajima,[3] Song Liu[8], Toshiya Okazaki,[9] Wei Chen,[6] Takeo Minari,[2,4,10] Yoshiki Sakuma,[10] Kazuhito Tsukagoshi,[2] Kazu Suenaga,[3] Takaaki Taniguchi,\*[2] and Minoru Osada\*[2,11]*

1 International Center for Young Scientists (ICYS), National Institute for Materials Science (NIMS), Tsukuba 305-0044, Japan

2 WPI International Center for Materials Nanoarchitectonics (WPI-MANA), National Institute for Materials Science (NIMS), Tsukuba 305-0044, Japan

3 Nanomaterials Research Institute, National Institute of Advanced Industrial Science and Technology, AIST Central 5, Tsukuba 305-8564, Japan

4 Center for Functional Sensor & Actuator (CFSN), National Institute for Materials Science (NIMS), Tsukuba 305-0044, Japan

5 The School of Materials Science and Engineering, Zhengzhou University, Zhengzhou 450001, P. R. China

6 Centre for Advanced 2D Materials and Department of Physics, National University of Singapore, 117546, Singapore

7 Institute of Materials Research and Engineering (IMRE), Agency for Science, Technology and Research (A*STAR), 138634, Singapore

8 Institute of Chemical Biology and Nanomedicine (ICBN), College of Chemistry and Chemical Engineering, Hunan University, Changsha 410082, P. R. China

9 CNT-Application Research Center, National Institute of Advanced Industrial Science and Technology, AIST Central 5, Tsukuba 305-8564, Japan

10 Research Center for Functional Materials, National Institute for Materials Science (NIMS), Tsukuba 305-0044, Japan

11 Institute of Materials and Systems for Sustainability (iMaSS), Division of Materials Research, Nagoya University, Nagoya 464-8603, Japan

**Abstract**

**Vapor transportation is the core process in growing transition-metal dichalcogenides (TMDCs) by chemical vapor deposition (CVD). One inevitable problem is the spatial inhomogeneity of the vapors. The non-stoichiometric supply of transition-metal precursors and chalcogen leads to poor control in products' location, morphology, crystallinity, uniformity and batch to batch reproducibility. While vapor-liquid-solid (VLS) growth involves molten precursors (e.g., non-volatile $Na_2MoO_4$) at the growth temperatures higher than their melting points. The liquid $Na_2MoO_4$ can precipitate solid $MoS_2$ monolayers when saturated with sulfur vapor. Taking advantage of the VLS growth, we achieved three kinds of important achievements: (i) 4-inch-wafer-scale uniform growth of $MoS_2$ flakes on $SiO_2$/Si substrates, (ii) 2-inch-wafer-scale growth of continuous $MoS_2$ film with a grain size exceeding 100 μm on sapphire substrates, and (iii) pattern (site-controlled) growth of $MoS_2$ flakes and film. We clarified that the VLS growth thus pave the new way for the high-efficient, scalable synthesis of two-dimensional TMDC monolayers.**

**Keywords:** chemical vapor deposition, vapor-liquid-solid, two-dimensional materials, wafer-scale growth, patterned growth

**1. Introduction**

Layered transition-metal dichalcogenides (TMDCs) present great variety in properties, including metal, semiconductor, insulator, and superconductor, which can be determined by their chemical composition and crystal structure [1,2]. Among them, the group 6 TMDC monolayers, i.e., $MoX_2$ and $WX_2$ (X = S, Se, Te), are endowed with intriguing optical and electrical properties and have been intensively studied in recent years [3-5]. In order to utilize atomically thin TMDCs for the next-generation, flexible, light-weight and wearable electronic devices, the controllable growth of large-area continuous film and high-quality flakes at specific location are highly demanded [6-8].

The synthesis of two-dimensional (2D) TMDCs via standard chemical vapor deposition (CVD) involves the usage of chalcogen, transition-metal oxides/chlorides/organics as volatile precursors. TMDC thin flakes of high crystallinity can therefore be deposited on various substrates through the transportation and reaction of vaporized precursors at high temperature [9]. The typical CVD of 2D $MoS_2$ monolayers was conducted by using powder precursors through a vapor-solid-solid (VSS) mechanism (nominated as powder CVD), where $MoO_3$ and S powders were loaded in a thermal tube furnace, and $SiO_2$/Si substrates were placed above the $MoO_3$ powder on a crucible for monolayer $MoS_2$ deposition [10,11]. This powder CVD has been widely used for growing various types of TMDCs using chalcogen (e.g., S, Se and Te) and transition-metal oxides, (e.g., $MoO_3$, $WO_3$) due to its simplicity and versatility. However, the disadvantages of powder CVD are the poor reproducibility and the coexisted byproducts during the growing cycles which limits the yield ratio and the homogeneity [12].

The spatially inhomogeneous transition-metal oxide vapor generated in powder CVD is the main reason for the poor coverage of TMDC monolayers on a substrate. The synthesized TMDC monolayers tend to appear only at the margins of the deposits on substrate which reduces their usability [12]. To date, the synthesis of wafer-scale TMDC film was achieved by using powder CVD with low pressure and high temperature, atomic layer deposition, and metal-organic CVD [6,13,14]. One drawback is that the as-grown TMDC films show small grains with a size of few or tens micrometers. The large density of defective grain boundaries in these TMDC films can be problematic when they are utilized in electronic devices due to their lower carrier mobility and chemical stability [15].

On the other hand, precise site-controlled (patterned) growth is of vital importance to integrate functional nanomaterials into nanoelectronic and optoelectronic devices, but which is nearly impossible for the TMDC monolayers grown by powder CVD due to the uncontrollable random nucleation. Recently, Han et al. and Li et al. reported of using patterned Au seeds to grow monolayer $MoS_2$ flakes with well-controlled locations [7,8], however, the residual Au seeds are somehow difficult to be removed.

Instead of using high vapor pressure transition-metal oxides to grow TMDCs, the non-volatile alkali molybdate and tungstate, e.g. $Na_2Mo_2O_7$, $Na_2MoO_4$ and $Na_2WO_4$, possessing extremely low vapor pressure and a suitable melting points around 700 ºC [16], had been demonstrated to grow high quality TMDC 1D nanoribbons and 2D isolated flakes through vapor-liquid-solid (VLS) process [17,18]. Here, in this paper, we further demonstrate the power of atmospheric pressure VLS growth in the following aspects: (i) uniform growth of monolayer $MoS_2$ flakes on 4-inch $SiO_2$/Si wafers, (ii) growth of continuous $MoS_2$ film with a grain size exceeding 100 μm on 2-inch sapphire substrates, (iii) patterned (site-controlled) growth of $MoS_2$ film and flakes without using Au seeds, (iv) high reproducibility in multi-batch growth, (v) universal mechanism for growing a large group of 2D TMDC crystals.

## 2. Results and discussion

### 2.1. Mechanism of VLS-growth of TMDC monolayers

The steps of wafer-scale and patterned growth of TMDC monolayers with non-volatile precursors via a VLS method is illustrated in **Fig. 1**. Taking VLS-MoS$_2$ monolayers as an example, Na$_2$MoO$_4$ particles are firstly dispersed on the growth substrate by spin-coating of its aqueous solution (**Fig. 1a**). When the growth temperature is higher than its melting point (687 $^o$C) in the CVD process, solid Na$_2$MoO$_4$ particles start to melt into liquid and wet the substrate surface. In the meanwhile, sulfur vapor begins to dissolve into the Na-Mo-O droplets (**Fig. 1b**). Finally, solid MoS$_2$ monolayers nucleate and grow up on the substrate from the Na-Mo-O-S liquid when it becomes sulfur-oversaturated (**Fig. 1c**). Alternatively, we can also pattern the Na$_2$MoO$_4$ particles on substrate with the aid of photolithography (**Fig. 1d**). After the sulfurization process, patterned MoS$_2$ monolayers with desired sites are grown on the substrate (**Fig. 1e**).

### 2.2. 4-inch wafer-scale VLS-growth of MoS$_2$ flakes on SiO$_2$/Si substrates

To begin with, we first demonstrate the growth of MoS$_2$ monolayers in a standard powder CVD as a reference, in which SiO$_2$/Si substrates were placed on top of a MoO$_3$-contained crucible. A clear variation of the deposited patterns in 3-batch growth (I, II and III in **Fig. 2a**) consisting of byproducts. oxides, thick and thin MoS$_2$ were all deposited on the SiO$_2$/Si substrates. The synthesized products in the power CVD are largely determined by the distance between the powder precursors and the substrates, which can be clearly distinguished from the color contrast of the grown specimen in **Fig. 2a** [12]. **Fig. 2b** shows an optical image of the specimen, in which the magnified images from three regions of interest (labeled c, d, e) are shown in **Fig. 2c-e**, respectively. Thick MoS$_2$ film with a mixture of MoO$_x$ were grown at the center of SiO$_2$/Si substrate (**Fig. 2c**), while MoS$_2$ monolayers were only grown at the margin (**Fig. 2d** and **2e**). It is worth to note that the surface coverage ratio of MoS$_2$ monolayers on the SiO$_2$/Si substrates is relatively low, ~1-5%. In the center of SiO$_2$/Si substrates, the highly concentrated transition metal cations penetrate deeply into the dielectric SiO$_2$ layer during the high-temperature powder CVD process which may lead to high leakage currents in field-effect transistors (FETs) [19]. This makes the following transfer process indispensable when evaluating the electrical properties of powder CVD grown TMDCs.

In sharp contrast to powder CVD, uniformly distributed monolayer MoS$_2$ flakes can be synthesized on a 4-inch SiO$_2$/Si wafer by using VLS-growth method (**Fig. 2f**). Firstly, Na$_2$MoO$_4$ particles were spin-coated (5000 rpm for 60 s of ~5 mg/mL Na$_2$MoO$_4$ aqueous solution) on a 4-inch SiO$_2$/Si wafer (**Fig. S1a-c**). The 4-inch silicon wafer was cut into 8 pieces with three different widths, ~15, 30 and 44 mm, in order to fit into a 2-inch tube furnace. Then, the substrates were loaded in the tube furnace as three layers stacking (**Fig. S1d** and **S1e**). The sulfurization was conducted at 750 °C for 5 min. **Fig. 2g** shows 8 typical optical images of MoS$_2$ flakes grown on the 8 pieces of SiO$_2$/Si substrates by using VLS growth. The distribution of the VLS-MoS$_2$ flakes on these SiO$_2$/Si substrates shows high uniformity with a lateral size of 5-40 μm and a surface coverage ratio of higher than 30%. Atomic force microscopy (AFM) image reveals the as-grown MoS$_2$ flakes are monolayers with a height of ~0.6 nm (**Fig. S1g**). The VLS-MoS$_2$ flakes show typical Raman modes at 384.9 ($E^1_{2g}$) and 404.6 cm$^{-1}$ ($A_{1g}$) and a sharp A-exciton peak (~1.84 eV) represents the direct optical transition of monolayer MoS$_2$, indicating the growth of 2H-phase MoS$_2$ (**Fig. S1h** and **S1i**) [4,20].

The fluorescence images of MoS$_2$ monolayers grown by powder CVD and the VLS method are shown in **Fig. 2h** and **2i**, respectively. The powder CVD grown MoS$_2$ shows nonuniform luminescence, bright at the center and weak at the margin of MoS$_2$ flakes (**Fig. 2h**). It is quite common to see this optical heterogeneity in the powder CVD grown TMDC monolayers, e.g., MoS$_2$, MoSe$_2$, WS$_2$ and WSe$_2$ [21-27]. We attribute the nonuniform luminescence to the structural deficiencies in these TMDC monolayers which may originate from the large spatial variation of transition-metal oxide vapor [27]. On the contrary, the VLS-MoS$_2$ flakes show homogeneous luminescence property (**Fig. 2i**), and we attribute this phenomenon to the simultaneous nucleation and growth of monolayer VLS-MoS$_2$ flakes from liquid Na$_2$MoO$_4$ on the whole substrate without employing nonuniform vapor transportation of Mo-containing species. **Fig. 2j** shows the statistical comparison of the PL peak distribution of the VLS-MoS$_2$ monolayers and the powder CVD grown MoS$_2$ flakes. One can see that the powder CVD grown MoS$_2$ flakes show highly diverse PL intensity and photon energy in sharp contrast to the homogeneous PL property of the VLS-MoS$_2$ monolayers (**Fig. S2**). It is worth to mention that the PL intensity is a widely accepted criteria when evaluating the quality of 2D TMDC monolayers. However, our recent studies based on optical (PL), electrical (transport property) and structural (atomic-resolution scanning electron transmission electroscope (STEM)) analysis have demonstrated that the structure defects and their interactions with adsorbed oxygen and water can greatly increase the PL intensity and degrade the carrier mobility of monolayer MoS$_2$ flakes [28]. The relative-uniform, low-intensity PL and narrow distribution of photon energy reveal the much more uniform optical quality and higher crystallinity of VLS-MoS$_2$ monolayers grown from Na$_2$MoO$_4$ than the monolayer MoS$_2$ flakes grown from powder MoO$_3$.

**Fig. 2k** shows the XPS spectra of VLS-MoS$_2$ flakes transferred onto an Au coated silicon substrate. The Mo$^{4+}$ 3d$_{5/2}$ and 3d$_{3/2}$ peaks are located at 229.5 eV and 232.7 eV, while the S$^{2-}$ 2p$_{3/2}$ and 2p$_{1/2}$ peaks are located at 162.2 eV and 163.5 eV, respectively. These results are consistent with the corresponding binding energies of Mo and S reported in bulk MoS$_2$ and CVD-grown MoS$_2$ monolayers [29,30]. Here, the VLS-MoS$_2$ flakes have a Mo/S atomic ratio around 0.5 corresponding to stoichiometric intrinsic MoS$_2$. One major concern of the VLS-MoS$_2$ monolayers is the doping/substitution of Na atoms in the MoS$_2$ lattice due to the usage of Na$_2$MoO$_4$. To fully understand the distribution of Na residuals, the as-grown VLS-MoS$_2$ flakes were transferred onto TEM grid directly without rinsing with IPA/H$_2$O (9/1 *vol.*) solution, was employed (**Fig. S3**). It is worth to note that the VLS-MoS$_2$ was found highly crystalline with very low concentration of S and Mo vacancies. As confirmed from the electron energy loss spectra of **Fig. S3c** and **S3d**, the Na residuals tend to form nanoclusters and absorb on the surface of the as-grown VLS-MoS$_2$ monolayers. The Na residuals can be removed easily by rinsing in IPA/H$_2$O (9/1 *vol.*) solution and pure IPA sequentially. As a result, no visible Na 1s signal at ~1071.5 eV was detected in the transferred VLS-MoS$_2$ monolayers (**Fig. 2k**).

**2.3. 2-inch wafer-scale VLS-growth of MoS$_2$ film on sapphire substrates**

High-quality and large-area TMDC film has been demonstrated for various practical applications in electronic devices [6,31,32]. The VLS growth of large-area MoS$_2$ film on sapphire substrate shows greater feasibility than on SiO$_2$/Si substrate due to the higher surface wettability of Na$_2$MoO$_4$ aqueous solution on sapphire substrate (**Fig. S4a-e**). The spin-coated

Na$_2$MoO$_4$ tend to spread over the surface homogeneously and form high area-density small particles on sapphire substrate as shown in **Fig. S4f**. **Fig. 3a** shows a photo of continuous VLS-MoS$_2$ film grown on a 2-inch sapphire substrate, appearing in light yellow-green color. **Fig. 3b** and **3c** are optical and SEM images of the VLS-MoS$_2$ film grown on a sapphire substrate. The VLS-MoS$_2$ film shows large grain size exceeding 100 μm in diameter. While in previous study, the grain sizes of wafer-scale TMDC film were often at the level of few or tens of micrometers [6,13,14,31]. Furthermore, the VLS-MoS$_2$ film has a surface coverage rate of ~99.6%, and only very small area of gaps and dots are existed at the grain boundaries. It is worth mentioning that the grain size of VLS-MoS$_2$ film can be controlled by tuning the quantity of sulfur vapor during the growth. **Fig. 3d** shows a transferred VLS-MoS$_2$ film on a SiO$_2$/Si substrate grown by using low concentrated sulfur (~80 mg sulfur for 30-minute growth). The less sulfur supply leads to low nucleation density of VLS-MoS$_2$ monolayers, as a result, the grain size can exceed 100 μm in diameter (**Fig. 3c** and **3d**). On the contrary, growing VLS-MoS$_2$ film with high concentrated sulfur (~60 mg sulfur for 10-minute growth) results in high nucleation density and smaller grain size (few-tens of micrometer in diameter) as shown in **Fig. 3f**. **Fig. 3e** and **3g** are corresponding fluorescence images of **Fig. 3d** and **3f**, respectively. The relatively uniform fluorescence indicates the VLS-MoS$_2$ film possessing high optical homogeneity. Furthermore, the narrow distribution of photon energy and PL intensity also demonstrate the uniform optical quality of the VLS-MoS$_2$ film (**Fig. 2j** and **Fig. S2).**

To investigate the electrical quality of the wafer-scale VLS-MoS$_2$ film, 78 FETs were fabricated and measured. All the FETs show high current on/off ratio in the range of $10^7$-$10^9$. The 31 FETs with grain boundary (GB) have an electron mobility of 9-30 cm$^2$/Vs and average electron mobility of 21.1 cm$^2$/Vs. While the 47 FETs without GBs have an electron mobility of 13-54 cm$^2$/Vs and average electron mobility of 24.8 cm$^2$/Vs. The similar average electron mobility of the two types VLS-MoS$_2$ FETs indicating the MoS$_2$ domains grown on the single-crystalline sapphire substrate may have good alignment and forming less defective GBs during the VLS growth.

**2.4. Patterned growth of VLS-MoS$_2$ film on sapphire substrates**

To integrate TMDCs with current printed electronics, it is of vital important to grow an array with sizable TMDC film without the use of photoresist. Here, we demonstrated a facile photomask- and vacuum ultraviolet (VUV)-mediated patterning method to selectively grow VLS-MoS$_2$ film on sapphire substrates. **Fig. 4a-e** schematically illustrate the procedures of selective deposition of Na$_2$MoO$_4$ particles on a sapphire substrate. In previous section, we have learned that the area-density and size of Na$_2$MoO$_4$ particles are highly related to the hydrophilicity of the growth substrate surface (**Fig. S4**). Here, we utilized a photomask with desired patterns to intimately contact with a sapphire substrate directly (**Fig. 4a**) and then exposed by VUV-irradiation (**Fig. 4b**). As a result, the VUV-exposed area became highly hydrophilic on the sapphire surface (**Fig. 4c**). After spin-coating with Na$_2$MoO$_4$ aqueous solution (**Fig. 4d** and **4e**), high area-density of small Na$_2$MoO$_4$ particles were deposited in the highly hydrophilic rectangles which was in sharp contrast to the non-VUV-exposed area (See **Fig. 4f-h**). In the non-VUV-exposed area, only sparse large Na$_2$MoO$_4$ particles were deposited and they failed to grow continuous film but thick MoS$_2$ islands. Therefore, an array of rectangular VLS-MoS$_2$ film was obtained by sulfurizing the patterned Na$_2$MoO$_4$ as illustrated in **Fig. 4i**. **Fig. 4j** shows a photo of the photoresist-free patterned growth of VLS-MoS$_2$ film (yellow-green color) on a sapphire substrate. The zoom-in optical image shown in **Fig. 4k** displays the high selectivity growth of rectangular VLS-MoS$_2$ film with a size of ~200×500 μm$^2$ at the VUV-exposed area. The VLS-MoS$_2$ film also shows a sharp boundary at the edge of rectangles indicating the well-defined pattern by our method (**Fig. 4l**). We propose this photoresist-free patterned growth method a great potential in application to flexible printed electronics [23,33].

### 2.5. Patterned growth of VLS-MoS$_2$ flakes on SiO$_2$/Si substrates.

Precise control the growth site of nanomaterials with high resolution is the core concern when integrating them into functional devices, such as logic circuits and photon emitters [32,34,35]. VLS growth had demonstrated a great capability of site-controlled of nanowires due to the formation of immobilized liquid precursors, which is almost impossible for the VSS growth in powder CVD. When patterning catalyst nanoparticles on substrates with unique crystal planes or well aligned surface steps, epitaxial VLS-growth of oriented nanowires or nanotubes were achieved [36]. Here, we explored the capability of growing high-quality VLS-MoS$_2$ flakes at desired locations via the VLS method without using Au seeds. **Fig. 5a** displays arrays of twelve ~40×40 μm$^2$ square windows on a SiO$_2$/Si substrate were fabricated by a standard photolithography. In our study, a layer of photoresist (AZ5214) was spin-coated on a SiO$_2$/Si substrate first. Then, a standard LED photolithography process was employed to fabricate square windows on the photoresist. Later, the patterned square windows of exposed SiO$_2$/Si substrate were dealt with a surface modification using UV-Ozone treatment and spin-coated with Na$_2$MoO$_4$ aqueous solution. Finally, the photoresist was removed through a lift-off process by rinsing with dehydrate acetone and IPA. The Na$_2$MoO$_4$ particles can therefore be deposited on the SiO$_2$/Si substrate at desired position and pattern (see **Fig. 5b** and inset). We noticed that most of Na$_2$MoO$_4$ particles were accumulated at the edge of the patterned squares. After the sulfurization of the patterned Na$_2$MoO$_4$ particles, monolayer VLS-MoS$_2$ flakes can be synthesized along the edge of the squares as shown in **Fig. 5c** and **5d**.

Due to the relatively lower growth temperature of VLS-MoS$_2$ flakes in the VLS method, we were able to fabricate FET devices right after the growth process and directly on the growth substrates without additional sample transferring. In this case, the intrinsic electronic properties of the as-grown monolayer VLS-MoS$_2$ flakes can be honestly presented. One can see that the VLS-MoS$_2$ flakes possess very small hysteresis in the transfer curve with high current on/off ratio (>10$^7$) (**Fig. 5e)** and high performance in the FET characteristic curve (**Fig. 5f**). The VLS-MoS$_2$ flakes demonstrate high electron mobility ranging from 9.2 to 52.4 cm$^2$/Vs based on a statistic of 52 FETs **(Fig. 5g)**. These provide clear evidence of great electrical performance and high crystallinity of the as-grown monolayer VLS-MoS$_2$ flakes.[35] Note that the leakage currents of the SiO$_2$ layer were retained in a few pA level which indicates the high quality dielectricity sustained after the CVD process (**Fig. S5**). We attribute the intact dielectric SiO$_2$ layer in the VLS growth process to the high efficiency of Na$_2$MoO$_4$ precursor and very small amount of precursor is required for the growth of MoS$_2$ monolayers.

## 3. Conclusions

In summary, we proposed and demonstrated the VLS growth of TMDC monolayers by using non-volatile precursors, e.g., Na$_2$MoO$_4$ for grown MoS$_2$ monolayers. The VLS growth also reveals great feasibility in large-scale, uniform growth of monolayer MoS$_2$ flakes and continuous film. Utilizing the unique nature of the non-volatile $Na_2MoO_4$, patterned (site-controlled) growth of VLS-$MoS_2$ film and flakes were also demonstrated. Furthermore, the non-volatile precursors hold great promise as ink materials for printing electric circuits. The VLS growth with the printed precursors presents great potential in fabricating sophisticated circuits which can be a great leap towards the utilize of TMDCs as core components in the future electronics.

**Electronic supplementary information (ESI)**

Supplementary data associated with this article can be found, in the online version, at doi:. The supplementary data include: (i) growth conditions for all the samples demonstrated in the manuscript, (ii) transfer of VLS-$MoS_2$ flakes for XPS and TEM study, (iii) characterization of VLS-$MoS_2$ monolayers: Raman, PL, STEM, EELS and FETs.

**Author information**

Corresponding Authors

S. Li: li.shisheng@nims.go.jp

T. Taniguchi: taniguchi.takaaki@nims.go.jp

M. Osada: mosada@imass.nagoya-u.ac.jp

**Author Contributions**

S. Li. designed and conducted the VLS growth. Y.-C. Lin. and K. Suenaga performed and interpreted the STEM data. X.-Y. Liu and T. Minari conducted the sapphire surface modification. Z. Hu and W. Chen performed the XPS study. J. Wu illustrated the VLS growth.

H. Nakajima and T. Okazaki conducted the Raman and PL study. S. Li, Y.-C. Lin and T. Taniguchi wrote the paper. All the authors discussed and commented on the manuscript.

**Conflicts of interest**

There are no conflicts of interest to declare

**Acknowledgment**

S.L. acknowledges the financial support from JSPS-KAKENHI (19K15399). T.T and M.O. acknowledge the financial support from JSPS-KAKENHI (17K19187). K.S. and Y.-C.L. acknowledge the support from JSPS-KAKENHI (JP16H06333) and (18K14119). W.C. acknowledges the financial support from Singapore MOE under grants of R143-000-652-112 and R143-000-A43-114. T. M. acknowledges the financial support from the MEXT of Japan under Grant-In-Aid for Scientific Research (No. 26286040 and 17H02769). S.L. acknowledges all staff members of the Nanofabrication group at NIMS and Dr. W. Ma for their support.

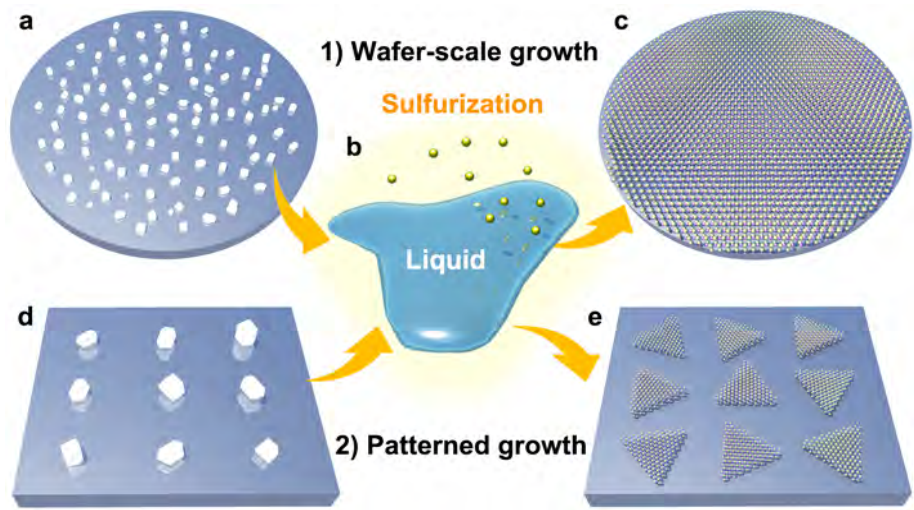

**Fig. 1.** Schematics of wafer-scale and patterned growth of MoS$_2$ monolayers via a VLS method. (a) Dispersed Na$_2$MoO$_4$ particles on sapphire wafer. (b) Molten Na$_2$MoO$_4$ droplet starts to wet the substrate surface. (c) Sulfurization of the molten Na$_2$MoO$_4$ droplets leads to the nucleation and grow up of VLS-MoS$_2$ film on the whole wafer. (d) Patterned Na$_2$MoO$_4$ particles on growth substrate with the aid of photolithography. (e) Patterned growth of monolayer VLS-MoS$_2$ flakes by sulfurization of the site-specific non-volatile Na$_2$MoO$_4$ droplets.

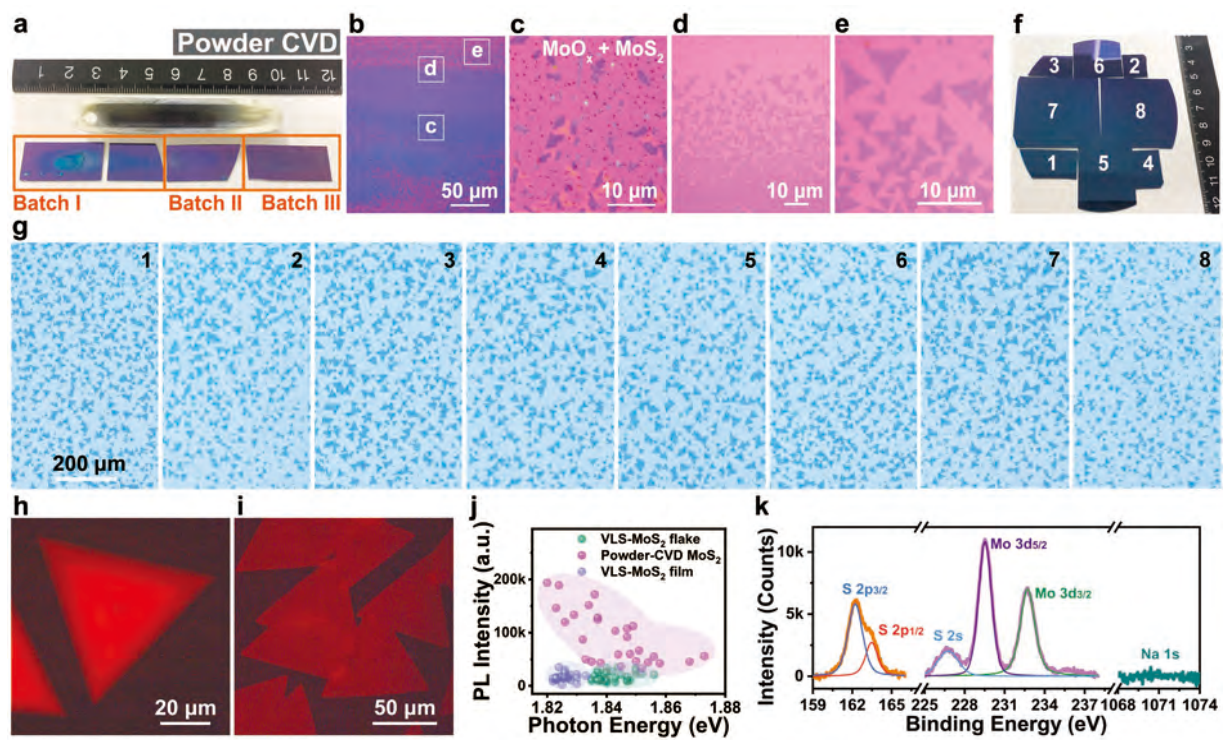

**Fig. 2.** Comparison of MoS$_2$ growth between powder CVD and VLS-growth methods. (a) A photo of 3-batch growth (I, II, III) of MoS$_2$ flakes using a powder CVD method. The color contrast indicates the nonuniformity of as-grown samples and poor reproducibility in 3-batch growth. (b) An optical image of the mixed products (MoS$_2$ and oxides) on growth substrate. (c-e) Detailed morphologies of the as-grown samples at three different positions labeled in (b). (f) A photo of 1-batch VLS growth of monolayer MoS$_2$ flakes on a 4-inch wafer. Substrate width: No. 1-4, ~15 mm; No. 5-6, ~30 mm; No. 7-8, ~44mm. (g) Typical optical images of monolayer VLS-MoS$_2$ flakes grown on the eight substrates labeled in (f). (h, i) Fluorescence image of monolayer MoS$_2$ flakes grown by (h) powder CVD and (i) VLS-growth methods. (j) PL distribution (photon energy and intensity) of three-type monolayer MoS$_2$ samples. (k) XPS spectra of transferred monolayer VLS-MoS$_2$ flakes on Au coated silicon substrates showing the fine spectra of Mo 3d, S 2p and Na 1s, respectively.

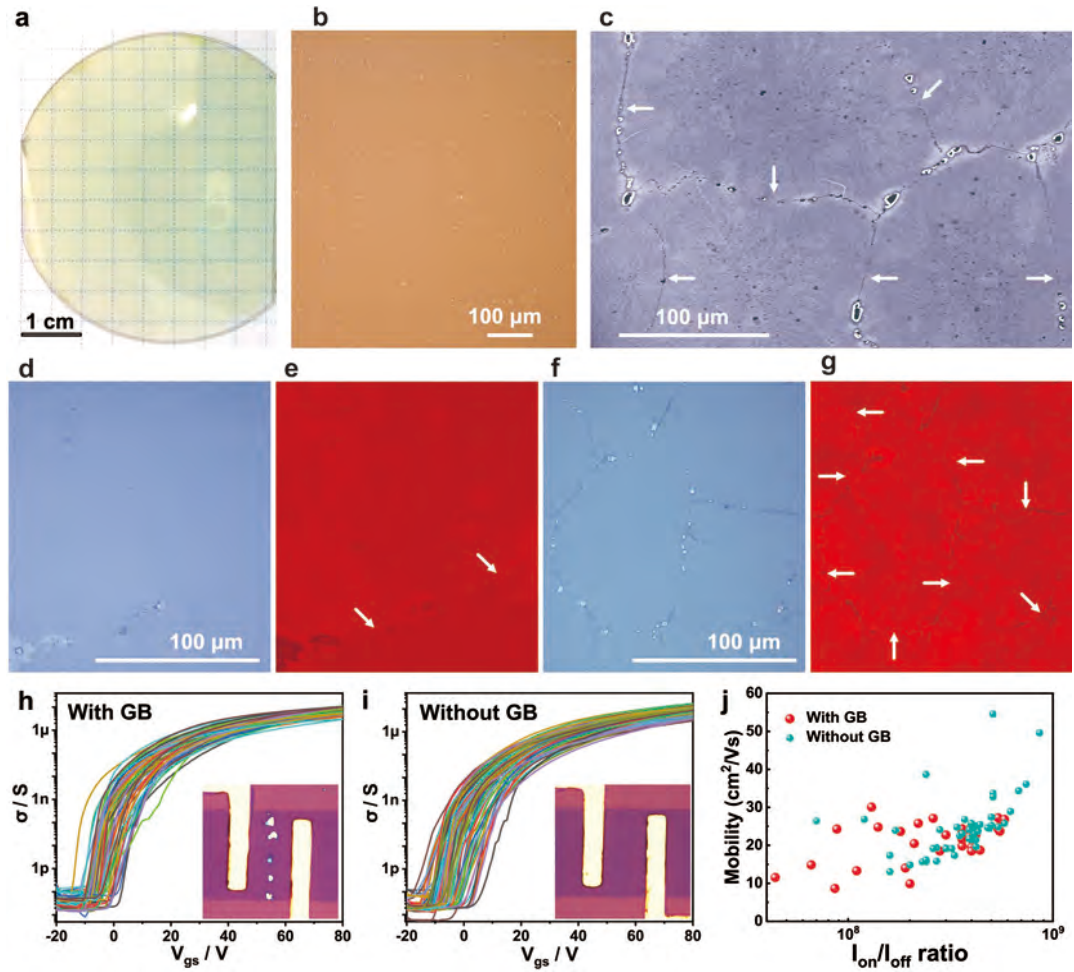

**Fig. 3.** 2-inch wafer-scale VLS-growth of MoS$_2$ film. (a) A photo of continuous monolayer VLS-MoS$_2$ film grown on a 2-inch sapphire substrate. (b, c) Typical (b) optical and (c) SEM image of the as-grown MoS$_2$ film. (d, e) Typical (d) optical and (e) fluorescence image of a VLS-MoS$_2$ film grown with low concentrated sulfur and long growth time. (f, g) Typical (f) optical and (g) fluorescence image of a VLS-MoS$_2$ film grown with high concentrated sulfur and short growth time. Figure (d)-(g) show the transferred VLS-MoS$_2$ film on two clean SiO$_2$/Si substrates. The arrows in (e) and (g) indicate the grain boundaries in VLS-MoS$_2$ films. (h, i) Transport curves of (h) 31 VLS-MoS$_2$ FETs with grain boundary (GB) and (i) 47 VLS-MoS$_2$ FETs without GB. (j) Distribution of electron mobilities and current on/off ratio of the 78 VLS-MoS$_2$ FETs.

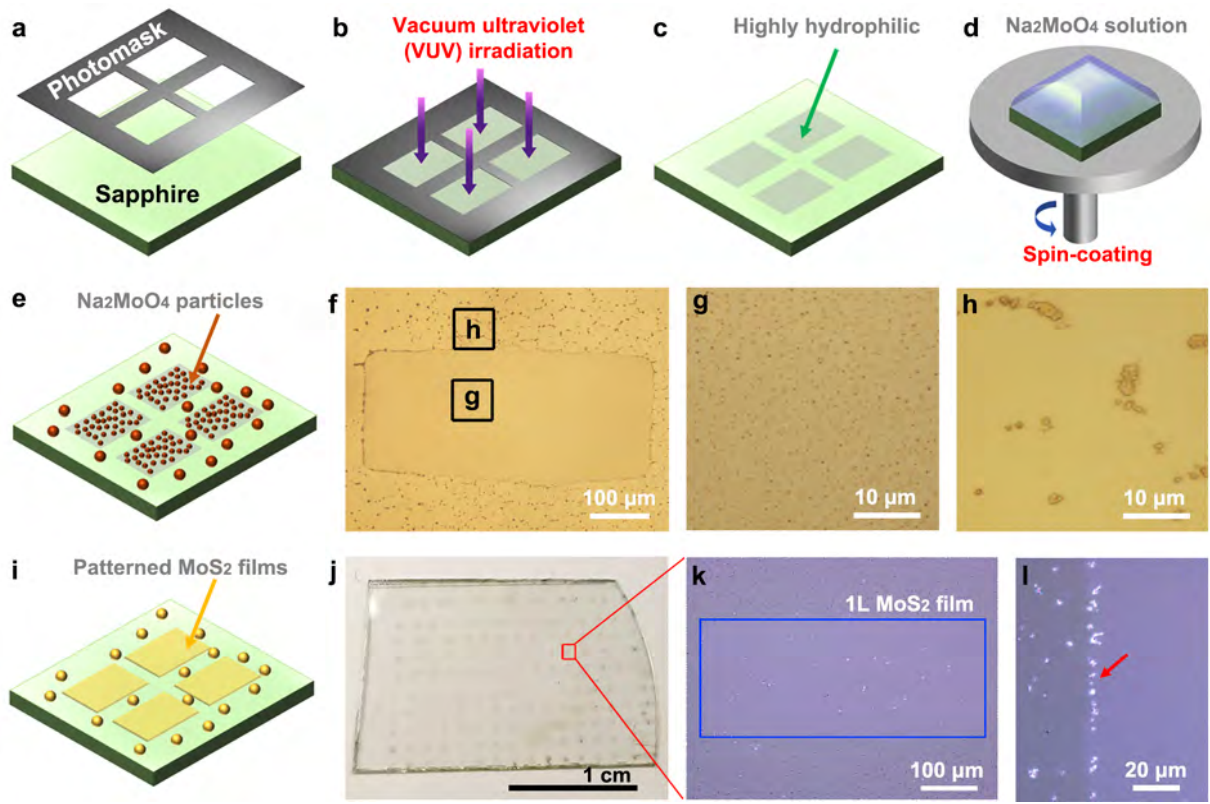

**Fig. 4. Patterned VLS-growth of monolayer MoS$_2$ film.** (a, b) Schematics of photomask mediated VUV-irradiation for patterning a sapphire substrate. (c) An array of highly hydrophilic rectangular area (the VUV-exposed area) formed on the sapphire substrate. (d) Spin-coating of Na$_2$MoO$_4$ aqueous solution on the patterned sapphire substrate. (e) Schematic of the patterned rectangular area with high area-density small Na$_2$MoO$_4$ particles on sapphire substrate. (f-h) Optical images of Na$_2$MoO$_4$ particles deposited on the patterned sapphire substrate. (g) An optical image shows high area-density and small Na$_2$MoO$_4$ particles were deposited in the highly hydrophilic rectangular area. (h) An optical image shows only sparse and large Na$_2$MoO$_4$ particles were deposited in the non-VUV-exposed area. (i, j) A schematic and a photo of patterned growth of VLS-MoS$_2$ film on a sapphire substrate. (k, l) Optical images of continuous VLS-MoS$_2$ film in the rectangular area (~200×500 μm$^2$) and the surrounding thick MoS$_2$ islands. The red arrow in (l) indicates the sharp edge of the rectangular VLS-MoS$_2$ film.

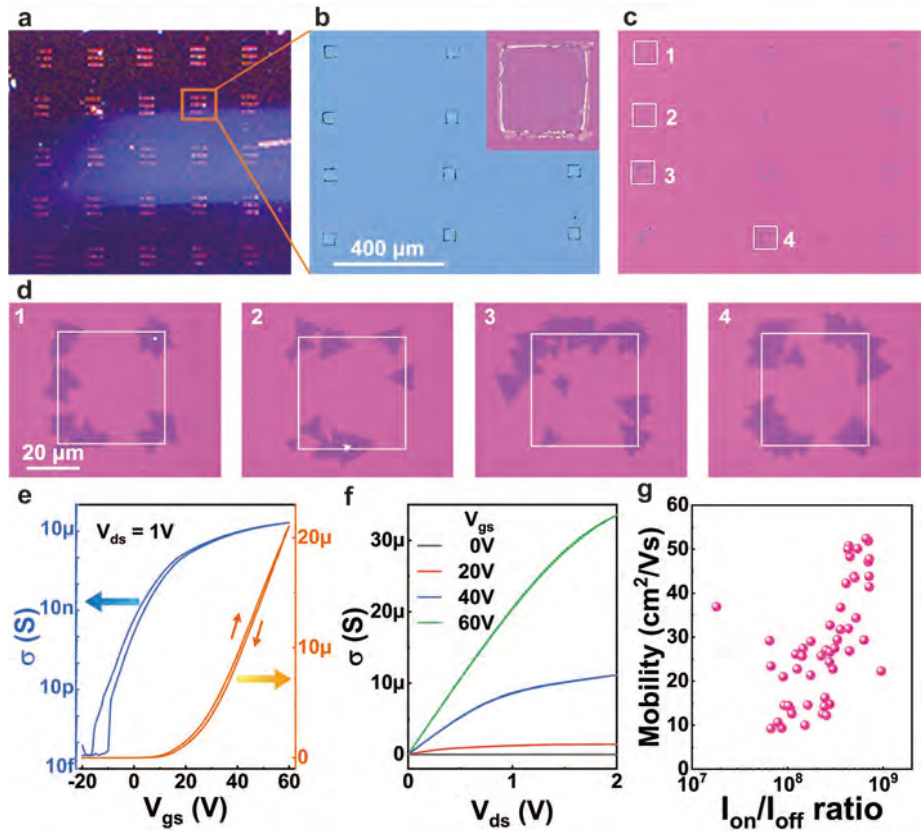

**Fig. 5.** Patterned VLS-growth of MoS$_2$ flakes. (a) Optical image of arrays of Na$_2$MoO$_4$ squares on a SiO$_2$/Si substrate. (b) Optical image of a unit cell with 12 squares (3×4). Inset is a enlarged view of a single square (~40×40 μm$^2$), the Na$_2$MoO$_4$ is accumulated at the edge. (c) Optical image of a unit cell of MoS$_2$ array. (d) Enlarged view of VLS-MoS$_2$ flakes grown from patterned Na$_2$MoO$_4$ squares. (e) Typical transport property of a VLS-MoS$_2$ FET. (f) Corresponding output performance of the VLS-MoS$_2$ FETs. (g) Distribution of electron mobility and current on/off ratio of 52 VLS-MoS$_2$ FETs.

# Wafer-Scale and Deterministic Patterned Growth of Monolayer MoS$_2$ via Vapor-Liquid-Solid Method

Shisheng Li,*[1,2] Yung-Chang Lin,[3] Xu-Ying Liu,[1,4,5] Zehua Hu,[6] Jing Wu,[7] Hideaki Nakajima,[3] Song Liu[8], Toshiya Okazaki,[9] Wei Chen,[6] Takeo Minari, [2,4,10] Yoshiki Sakuma,[10] Kazuhito Tsukagoshi,[2] Kazu Suenaga,[3] Takaaki Taniguchi,*[2] and Minoru Osada*[2,11]

1 International Center for Young Scientists (ICYS), National Institute for Materials Science, Tsukuba 305-0044, Japan

2 WPI International Center for Materials Nanoarchitectonics (WPI-MANA), National Institute for Materials Science , Tsukuba 305-0044, Japan

3 Nanomaterials Research Institute, National Institute of Advanced Industrial Science and Technology, AIST Central 5, Tsukuba 305-8564, Japan

4 Center for Functional Sensor & Actuator (CFSN), National Institute for Materials Science, Tsukuba 305-0044, Japan

5 The School of Materials Science and Engineering, Zhengzhou University, Zhengzhou 450001, P. R. China

6 Centre for Advanced 2D Materials and Department of Physics, National University of Singapore, 117546, Singapore

7 Institute of Materials Research and Engineering (IMRE), Agency for Science, Technology and Research (A*STAR), 138634, Singapore

8 Institute of Chemical Biology and Nanomedicine (ICBN), College of Chemistry and Chemical Engineering, Hunan University, Changsha 410082, P. R. China

9 CNT-Application Research Center, National Institute of Advanced Industrial Science and Technology, AIST Central 5, Tsukuba 305-8564, Japan

10 Research Center for Functional Materials, National Institute for Materials Science, Tsukuba 305-0044, Japan

11 Institute of Materials and Systems for Sustainability (iMaSS), Division of Materials Research, Nagoya University, Nagoya 464-8603, Japan

**Growth conditions:**

*1) 4-inch-wafer-scale growth of monolayer VLS-MoS$_2$ flakes:* A 4-inch SiO$_2$/Si wafer was treated with UV-Ozone for 30 minutes to obtain hydrophilic surface. To load Na$_2$MoO$_4$, ~10 mL of 4 mg/mL Na$_2$MoO$_4$ aqueous solution was drop on the wafer first. Then, spin-coating was conducted at 5000 rpm for 60 seconds. The 4-inch SiO$_2$/Si wafer was cut to small pieces to fit the inner-diameter of a 2-inch quartz tube. The growth was performed in a thermal tube furnace at 750 °C for 5 minutes with 100 sccm high-purity Argon as carrier gas. The temperature ramping rate was 30 °C/min. The temperature of sulfur was kept at ~170 °C and ~25 mg sulfur was consumed during the growth. (**Fig. S1**)

*2) 2-inch-wafer-scale growth of monolayer VLS-MoS$_2$ film:* 2-inch sapphire substrate was pretreated with UV-Ozone for 30 minutes to obtain highly hydrophilic surface. Here, Na$_2$MoO$_4$ particles was uniformly distributed on sapphire substrate by loading ~1.5 mL of 5 mg/mL Na$_2$MoO$_4$ aqueous solution. And then, spin-coating was conducted at 3000 rpm for 60 seconds. The growth condition is like the growth of VLS-MoS$_2$ flakes on SiO$_2$/Si substrates. To engineer the grain size of as-grown VLS-MoS$_2$ film, the supply of sulfur and growth time were modified. For large-grain VLS-MoS$_2$ film (>100 μm), low concentrated sulfur and long growth time were used, ~80 mg sulfur was consumed in 30-minute growth. For small-grain VLS-MoS$_2$ film (<100 μm), high concentrated sulfur and short growth time were required, ~60 mg sulfur was consumed in 10-minute growth.

*3) Patterned growth of monolayer VLS-MoS$_2$ film:* A photomask was used to intimately contact with a sapphire substrate first. Then vacuum ultraviolet (VUV) was employed to irradiate the sapphire substrate for 2 minutes. As a result, the originally less-hydrophobic sapphire substrate with a contact angle of ~26.3° with 5 mg/mL Na$_2$MoO$_4$ aqueous solution was patterned with an array of highly hydrophilic rectangles. The VUV-exposed sapphire substrate usually has a highly hydrophilic surface with a contact angle of ~8.5°. Then, ~0.5 mL of 5 mg/mL Na$_2$MoO$_4$ was loaded on the sapphire substrate and spin-coated with a speed of 4000 rpm for 30 seconds. A clear difference of the size and area-density of Na$_2$MoO$_4$ particles between the highly hydrophilic rectangular area and surrounding less-hydrophilic area can be observed.

The growth results indicated that high area-density of small $Na_2MoO_4$ particles in the hydrophilic rectangular area were facile for the growth of continuous VLS-$MoS_2$ film. While the large $Na_2MoO_4$ particles deposited on the surrounding hydrophobic area were only converted to thick $MoS_2$ islands. (**Fig. 4**)

*4) Patterned growth of monolayer VLS-$MoS_2$ flakes:* A layer of photoresist (AZ5214) was spin-coated on a $SiO_2$/Si substrate. A standard LED photolithography process was performed to prepare patterns on the $SiO_2$/Si substrate. Then, the $SiO_2$/Si substrate was treated with UV-Ozone for 30 minutes to modify the surface to hydrophilic. After spin-coating of ~0.5 mL of 5 mg/mL $Na_2MoO_4$ aqueous solution at 3000 rpm for 30 seconds. The patterned $Na_2MoO_4$ was obtained by sequential washing with dehydrate acetone at 50 °C for 30 minutes and IPA for 5 minutes. Then the $SiO_2$/Si substrate with patterned $Na_2MoO_4$ particles was blew dry with $N_2$. The growth condition was the same as the growth of VLS-$MoS_2$ flakes on $SiO_2$/Si substrates. (**Fig. 5**)

**Transfer of VLS-$MoS_2$ flakes:**

*1) For XPS:* To remove the adsorbed Na residuals, the as-grown VLS-$MoS_2$ flakes were immersed in IPA/$H_2O$ (9/1 *vol.*) solution for 30 minutes. Then, the as-grown samples were spin-coated with a layer of PMMA. The PMMA film was peeled off from growth substrate using KOH (35 *wt*%) as etchant. The PMMA film was transferred to an Au (50 nm) coated heavily doped silicon substrate. Finally, the PMMA was removed by resining with acetone and IPA sequentially.

*2) For TEM observation:* To investigate the Na residuals in the as-grown VLS-$MoS_2$ flakes on $SiO_2$/Si substrate, the samples were firstly protected by spin-coating a layer of polycarbonate. The $SiO_2$ layer was therefore etched in a diluted HF solution and the polycarbonate along with the $MoS_2$ monolayers were peeled off from the growth substrate. The film was rinsed for several times in DI water before transferring to TEM grids. Finally, the polycarbonate film was removed in chloroform and the as-grown VLS-$MoS_2$ flakes left on a Quantifoil microgrids. (**Fig. S3**)

**Fabrication of VLS-MoS$_2$ FETs:**

The VLS-MoS$_2$ monolayers were grown on silicon substrates with 285-nm-thick SiO$_2$ layer. First, the substrates were spin-coated with a layer of photoresist (AZ5214). Then, a standard LED photolithography process was conducted to define the patterns of electrodes. After the deposition of Ti/Au (5 nm/50 nm) film, a lift-off process was employed to remove residual photoresist. A second photolithography and oxygen plasma etching were used to define the shape of the channels VLS-MoS$_2$ FETs. (**Fig. S5**)

**Characterization:**

*1) Raman and PL:* The micro-Raman/PL was performed using a laser confocal microscope (inVia, Renishaw). The 532-nm excitation laser was focused on the sample surface with a 100× objective lens. Then, Raman/PL signals from the MoS$_2$ samples were detected by an electron multiplying CCD detector (Andor) through a grating with 1800 grooves/mm for Raman and 150 grooves/mm for PL. The laser spot size was about 1 μm in diameter.

*2) STEM and EELS:* STEM images were acquired by using JEOL 2100F microscope equipped with dodecaple correctors and the cold field emission gun operating at 60 kV. The probe current was about 25-30 pA. The convergence semiangle was 35 mrad and the inner acquisition semiangle was 79 mrad. The EELS core loss spectra were taken by using Gatan low-voltage quantum spectrometer.

*3) Measurement of VLS-MoS$_2$ FETs:* To test the transport properties of VLS-MoS$_2$ FETs, the devices were loaded in a vacuum chamber with a pressure of ~10$^{-3}$ Pa. The backgate bias ($V_{gs}$) was scanned forward from -20 V to 60 V and backward from 60 V to -20 V, the source-drain bias ($V_{ds}$) is 1 V. To measure the output performance of VLS-MoS$_2$ FETs, the backgate bias ($V_{gs}$) was scanned from 0 to 60 V with a step of 20 V. The source-drain bias ($V_{ds}$) was swept forward from 0 V to 2 V and backward form 2 V to 0 V with a step of 50 mV. The Y-axis was normalized to conductivity using the channel current and device dimensions.

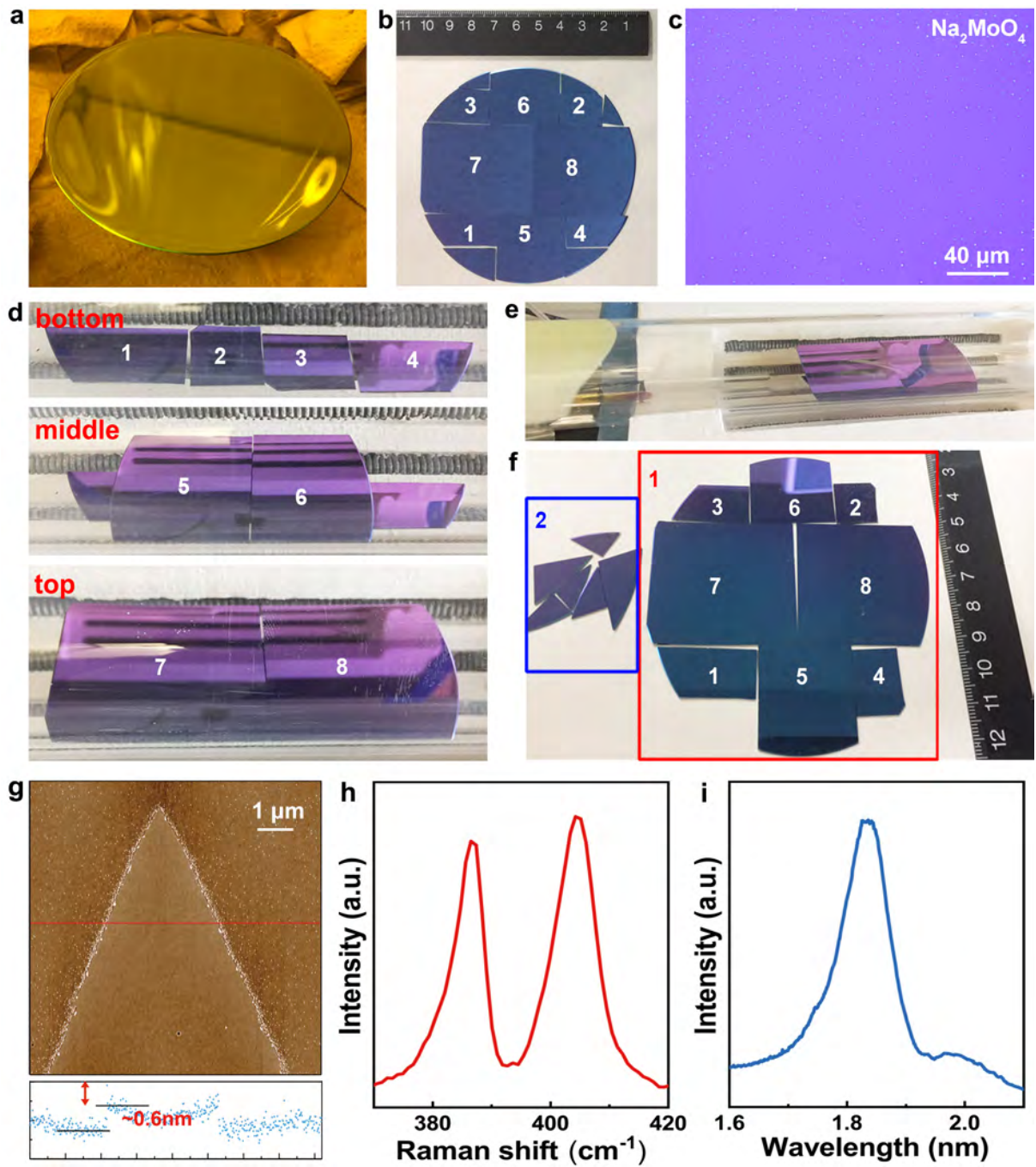

**Fig. S1. 4-inch-wafer-scale growth of monolayer VLS-MoS$_2$ flakes.** (a) Spin-coating of Na$_2$MoO$_4$ aqueous solution on a 4-inch SiO$_2$/Si wafer. (b) A photo of the SiO$_2$/Si substrates cut with different widths (~15, 30 and 44 mm). (c) An optical image of Na$_2$MoO$_4$ particles deposited on the SiO$_2$/Si substrate. (d) Photos of the SiO$_2$/Si substrates loaded in a 2-inch quartz tube furnace as three layers. (e) A photo of the tube furnace after VLS growth of MoS$_2$. (f) A photo of the 8 pieces of SiO$_2$/Si substrates after VLS growth of MoS$_2$. (g) A typical AFM image of a VLS-MoS$_2$ flake with a height of ~0.6 nm. (h, i) Typical (h) Raman and (i) PL spectra of the as-gown monolayer VLS-MoS$_2$ flakes.

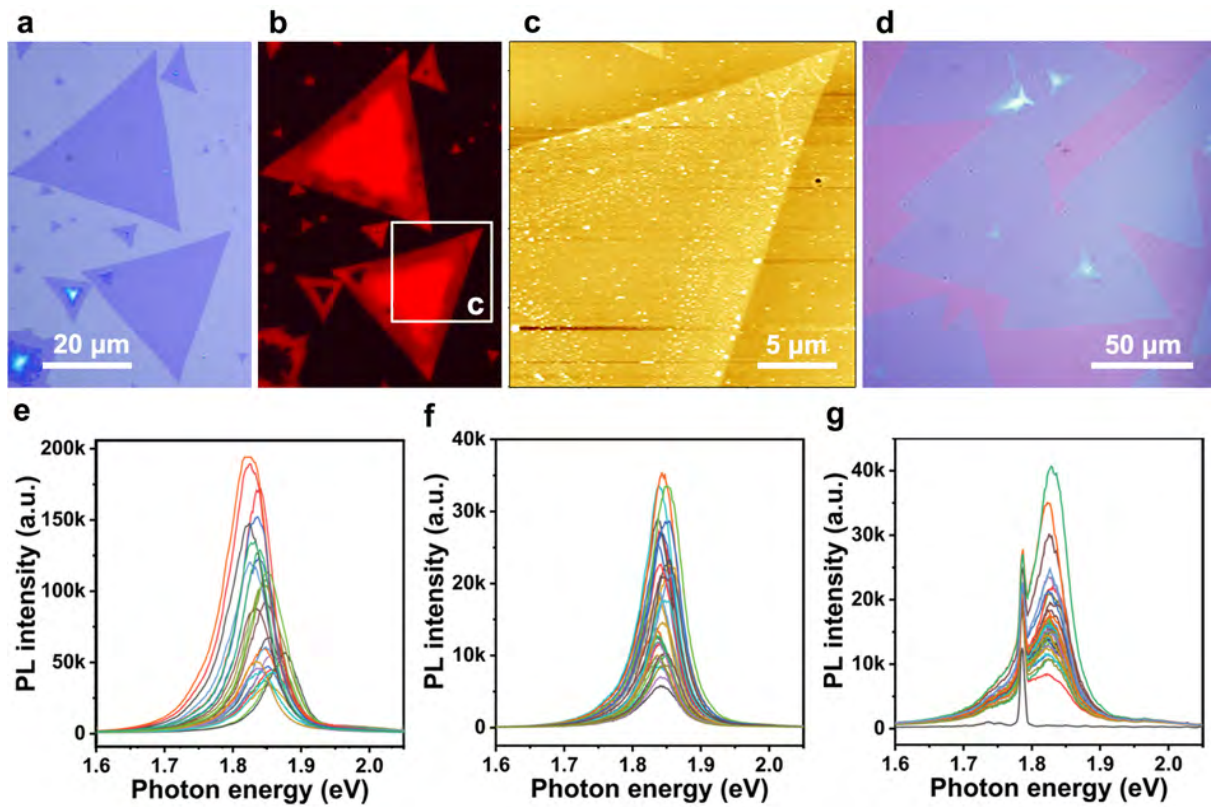

**Fig. S2** (a) An optical image and (b) corresponding fluorescence image of MoS$_2$ flakes on a SiO$_2$/Si substrate grown by powder CVD. (c) An AFM image shows the MoS$_2$ flake with non-uniform fluorescence is monolayer. (d) An optical image of monolayer VLS-MoS$_2$ flakes (corresponding to **Fig. 2i**). (e) PL spectra of monolayer MoS$_2$ flakes on a SiO$_2$/Si substrate grown by powder CVD. (f) PL spectra of monolayer VLS-MoS$_2$ flakes grown on a SiO$_2$/Si substrate. (g) PL spectra of monolayer VLS-MoS$_2$ film grown on a 2-inch sapphire substrate. The peak at ~1.78 eV comes from the sapphire substrate.

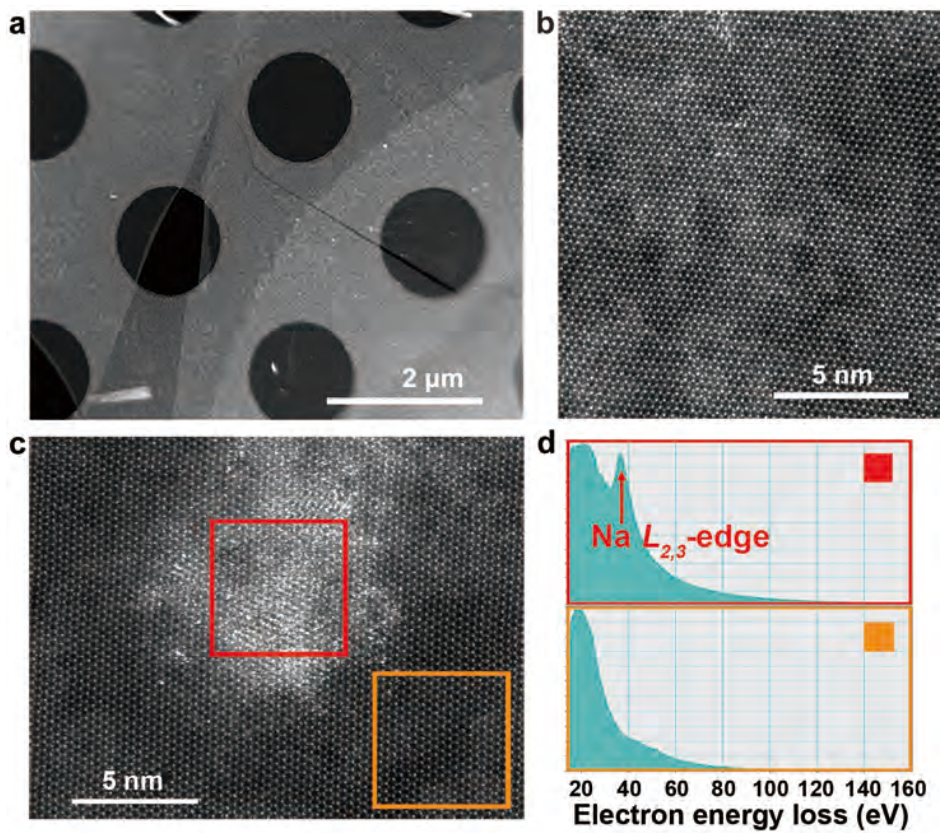

**Fig. S3. STEM images and EELS spectra of monolayer VLS-MoS$_2$ flakes.** (a) A low-magnification STEM image of a VLS-MoS$_2$ flake transferred on a Quantifoil microgrid. (b) A high-magnification STEM image of a VLS-MoS$_2$ flake with a high crystallinity. (c) A high-magnification STEM image of a VLS-MoS$_2$ flake with residual nanoparticle on surface. (d) EELS spectra taking from the residual nanoparticle (red square in (c)) and the clean area (orange square in (c)), respectively. The residual nanoparticle contains Na and shows EELS $L_{2,3}$-edge at ~35 eV.

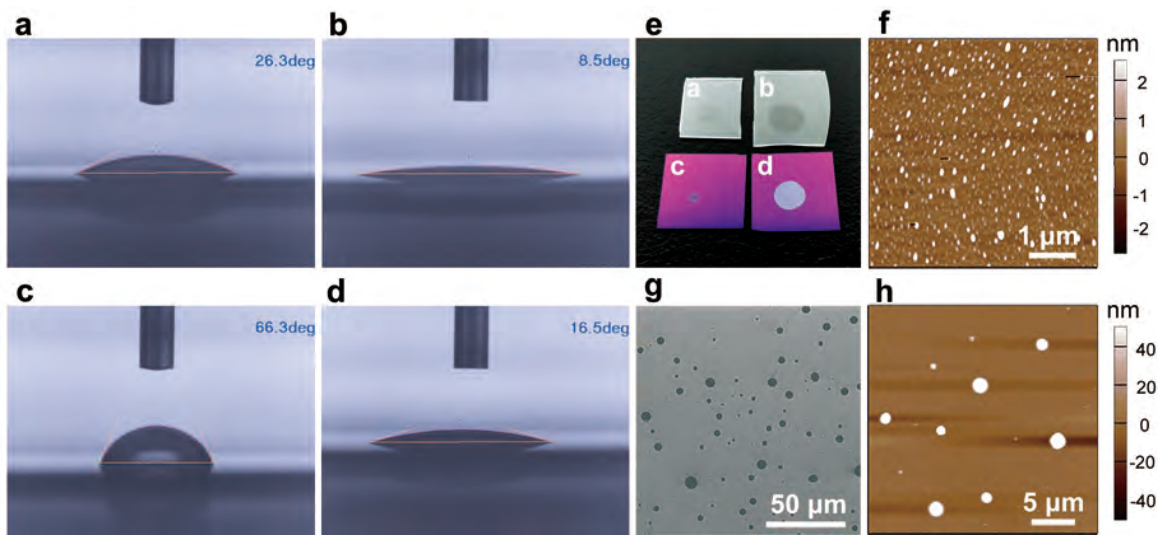

**Fig. S4. Contact angles of Na$_2$MoO$_4$ aqueous solution and as-deposited Na$_2$MoO$_4$ particles on SiO$_2$/Si and sapphire substrates, respectively.** (a, b) Optical images showing the contact angles of 2 μL of 5 mg/mL Na$_2$MoO$_4$ aqueous solution on (a) original and (b) UV-Ozone treated sapphire substrates. (c, d) Optical images showing the contact angles of Na$_2$MoO$_4$ aqueous solution on (c) original and (d) UV-Ozone treated SiO$_2$/Si substrates. (e) A photo showing the dried Na$_2$MoO$_4$ marks on the four substrates corresponding to **Fig. S4a-d**. (f) An AFM image of spin-coated Na$_2$MoO$_4$ particles on a sapphire substrate. (g) A SEM and (h) an AFM image of spin coated Na$_2$MoO$_4$ particles on a SiO$_2$/Si substrate. Substrates in **Fig. S4b** and **S4d** were treated with UV-Ozone for 30 minutes. For **Fig. S4f-h**, 5 mg/mL Na$_2$MoO$_4$ aqueous solution was used for spin-coating at a speed of 3000 rpm for 60 seconds.

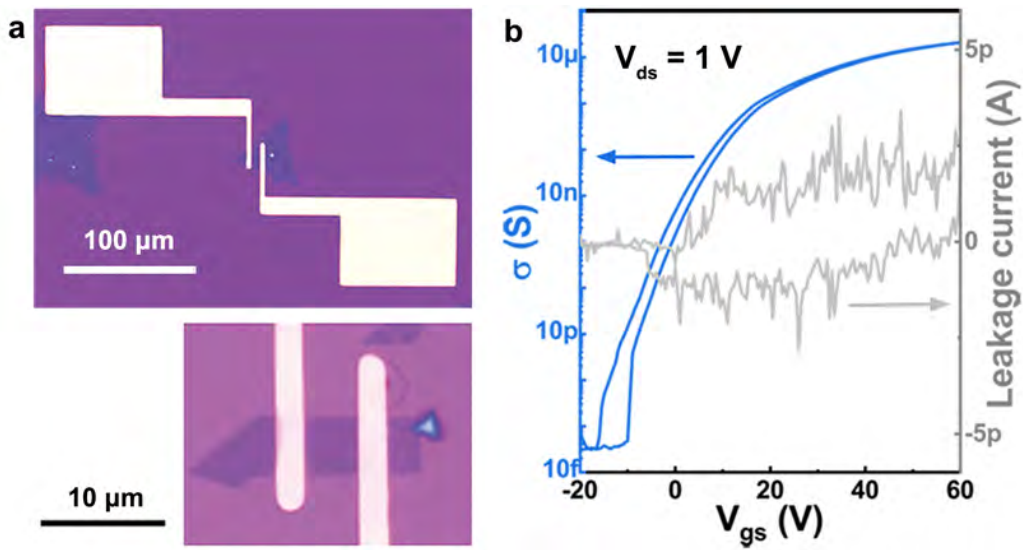

**Fig. S5. VLS-MoS$_2$ FETs.** (a) Optical images of an as-fabricated VLS-MoS$_2$ FET. (b) Typical transfer curve and leakage current of the VLS-MoS$_2$ FET shown in **Fig. 5e**.